\documentclass[aps,twocolumn,amssymb,floatfix,preprintnumbers,amsmath,amssymb,superscriptaddress]{revtex4}
\usepackage{graphicx}%
\usepackage[final]{pdfpages}
\usepackage{epstopdf}
\relax
 \pdfoutput=1

\begin{document}

\title{Interface size dependence of the
Josephson critical behaviour in pyrolytic graphite TEM lamellae}
\author{Ana Ballestar}
\affiliation{Division of Superconductivity and Magnetism, Institut
f\"ur Experimentelle Physik II, Universit\"{a}t Leipzig,
Linn\'{e}stra{\ss}e 5, D-04103 Leipzig, Germany}
\author{Tero T. Heikkil\"a}
\affiliation{Department of Physics and Nanoscience Center, University
  of Jyv\"askyl\"a,
    P.O. Box 35 (YFL),
    FI-40014 University of Jyv\"askyl\"a, Finland}
\author{Pablo Esquinazi}\email{esquin@physik.uni-leipzig.de}
\affiliation{Division of Superconductivity and Magnetism, Institut
f\"ur Experimentelle Physik II, Universit\"{a}t Leipzig,
Linn\'{e}stra{\ss}e 5, D-04103 Leipzig, Germany}
\begin{abstract}
We have studied the transport properties of TEM lamellae obtained from
a pyrolytic graphite sample with electrical contacts at the edges
of the embedded interfaces. The temperature dependence of the
resistance as well as the current-voltage characteristic curves
are compatible with the existence of Josephson coupled
superconducting regions. The transition temperature at which the
Josephson behavior sets in, decreases with the interface
width and vanishes for width below 200~nm.  This unexpected
behavior is apparently due to the influence
of weak localization effects on the superconducting critical
temperature.

\end{abstract}

\maketitle
\bigskip


The properties of surfaces and interfaces in solids can be
fundamentally different from those of the corresponding bulk
material. Differences in the carrier dispersion relation compared to the quadratic free electron case can strongly affect the electronic
properties of the interfaces. Conductivity experiments are extremely sensitive to such effect, specially in samples where superconductivity at certain interfaces compete with the rest of
the nonsuperconducting matrix \cite{rey07,goz08}. The possibility
of high-temperature superconductivity at surfaces and interfaces
has attracted the attention of the physics
community since the earliest 60's \cite{gin64}. Superconductivity
has been found at the interfaces between oxide insulators
\cite{rey07} and between metallic and insulating copper
oxides  with critical temperature $T_c \gtrsim 50~$K
\cite{goz08}. Interfaces in pure \cite{gip92,mun95,mun06,mun07} as
well as  doped Bi bicrystals can show superconductivity up to $T_c
\simeq 21~$K \cite{mun08}, although Bi as bulk is not a
superconductor.

High-temperature
superconductivity has been predicted to occur at topologically
protected flat bands on the surface \cite{kop11} or at certain
rhombohedral-Bernal interfaces \cite{kop13,mun13} in graphite
\cite{vol13}. Inclusions
of rhombohedral graphite ordered regions have been found embedded
in bulk Bernal graphite \cite{lin12} as well as in exfoliated
multilayer graphene films \cite{lui11}, both taken from highly oriented pyrolytic graphite (HOPG) samples.
The possible existence of high-temperature
superconductivity embedded in disordered graphite  has already been speculated
40~years ago  \cite{ant74,ant75}. Since then different studies were published in graphite
\cite{esqpip} as well as in doped disordered carbon samples \cite{fel14} 
providing indications of high transition temperatures.
The existence of quasi two-dimensional (2D) interfaces in  HOPG and
in Kish graphite samples has been known long time ago \cite{ina00},
but their extraordinary properties were
reported only recently as a result of contacting
the edges of the embedded interfaces found in
transmission electron microscope (TEM) graphite lamellae \cite{bal13}.
Evidence for
granular superconductivity in graphite flakes has been
independently observed in  the field hysteresis and field
periodic oscillations in high resolution magnetoresistance
measurements  as well as in SQUID magnetization
measurements of HOPG
samples with interfaces, for a review see \cite{esqpip}.

 Sample lateral-size dependent effects in the
magnetoresistance of graphite have been already reported and
explained in terms of the large (several microns) mean free path
of the carriers within the graphene layers \cite{gon07,dus11}. The
phenomenon we describe in this letter, however, has not been
reported yet for graphite and it is related to the transition
temperature $T_c$ characteristic of the Josephson effect found at
graphite interfaces \cite{bal13}. We have found that this $T_c$ decreases with
the width of the 2D interface region. The observed behavior may
clarify the origin of certain differences in the temperature
dependence of the resistance, especially the temperature of the
maximum $T_{\rm max}$ in the resistance of bulk graphite samples
and of small graphite flakes with interfaces.  We propose that, at least part of  this
difference is related to the area of the embedded interfaces where
superconductivity exists. We expect that there exists a direct correspondence
between the true superconducting critical temperature and the
temperature $T_c$ (at which the Josephson behavior starts to be
measurable)  and that $T_{\rm max} \sim T_c$.

\begin{figure}[tbp]\centering
\includegraphics[width=0.9\columnwidth]{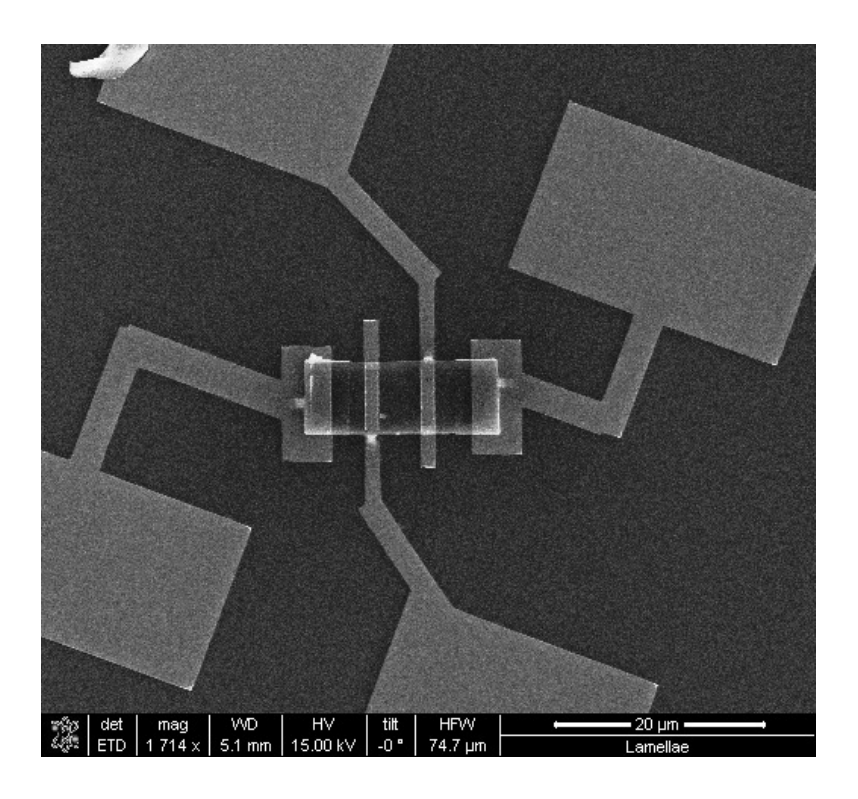}
\caption{Scanning electron microscope image of  one of the studied lamella
(thickness 200~nm).
The brighter areas correspond to the four Pt/Au electrodes 
placed in line as usual in the four probes method. Note that the outer 
contacts (used for the input current) are placed at the edges of the sample, i.e. at the edges of the interfaces. The inner ones are used to measure the corresponding response (voltage) coming from these interfaces.
The $c$-axis of the graphite structure (normal to the
graphene planes) is
parallel to the substrate, normal to the input current direction.}
\label{photo}
\end{figure}

TEM lamellae, with thickness between 80~nm and 800~nm, were cut
with the Ga$^+$ beam of a dual beam microscope (FEI Nanolab XT
200) from the same bulk HOPG sample grade ZYA ($0.4^\circ$ rocking
curve width). We previously covered the sample surface with a wolfram-carbide (WC)
film, deposited using Electron Beam Induced Deposition (EBID), to
avoid the penetration of the Ga$^+$ ions in the graphite structure
of the lamella \cite{bal13}. The lamella thickness corresponds to
the width of the graphene planes and therefore to the width of the
internal interfaces, as they run parallel to the graphene planes, see TEM images in
\cite{ina00,bal13}. Moreover, electron backscattering diffraction (EBSD) measurements showed the existence of grains of several micrometers in the $ab$-plane \cite{ina00,gon07}. The transport
measurements were done using a four-probe electrode technique, in
the conventional configuration, see Fig.~\ref{photo}.
\begin{figure}[tbp]
\centering
\includegraphics[width=1.1\columnwidth]{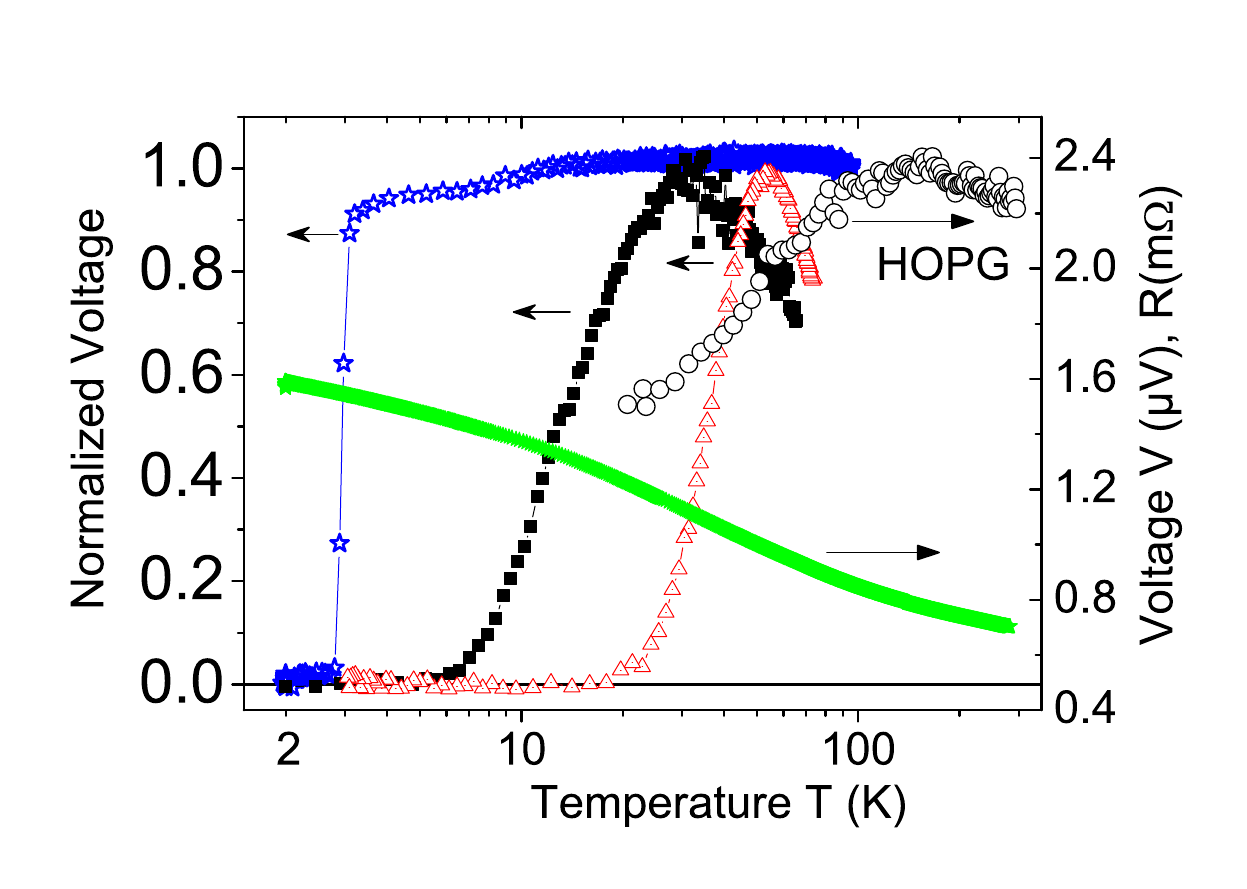}
\caption{Left $y$-axis: normalized voltage measured at 1~nA
current vs. temperature of three lamellae with different
thickness, namely 200, 300 and 500~nm ($\pm 20~$nm) for ($\star,
\blacksquare, \triangle$) and different normalization voltages, respectively.  
Right $y$-axis: Green symbols $(\ast)$ show the absolute
voltage vs. temperature at 1~nA measured for a lamella with
thickness $100 \pm 10$~nm. ($\circ$): Resistance $R$ vs.
temperature measured for the original bulk HOPG sample (current $1~\mu$A) from where
the lamellae were obtained.} \label{VT}
\end{figure}

Figure~\ref{VT} shows the temperature dependence of the voltage
for four TEM lamellae  measured at a constant input current of
1~nA. For clarity and  taking into account
the differences in the absolute resistance between the samples,
the voltages of three lamellae are normalised at their maxima (left $y$-axis).
The absolute resistance of a bulk HOPG samples
as well as the voltage of a fourth lamella are shown in the same figure (right $y$-axis).
Note that the measurements of the four lamellae presented in Fig.~\ref{VT}
were done using 1~nA  current because  the observed transitions
 shift to lower temperatures the larger the input
current, as expected in a system with superconducting grains.

\begin{figure}[tbp]\centering
\includegraphics[width=1.15\columnwidth]{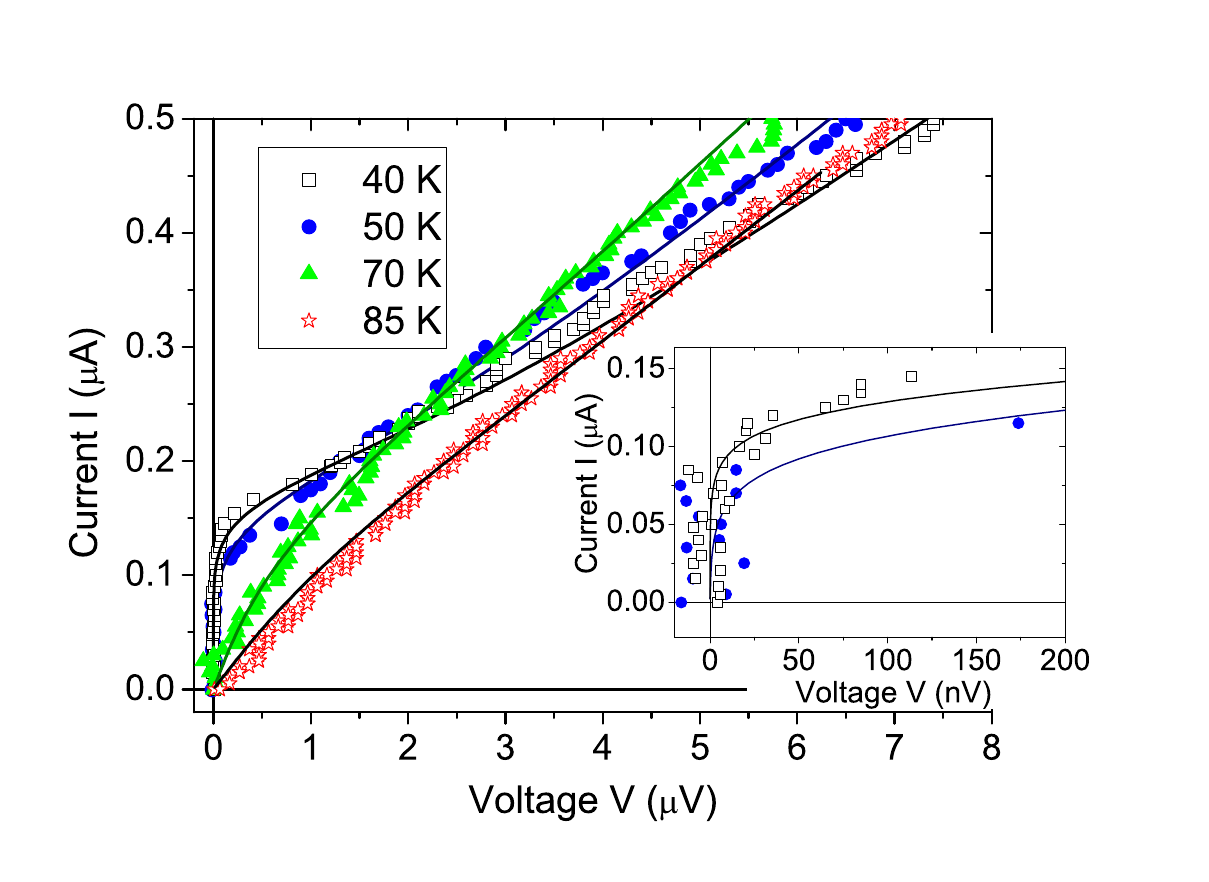}
\caption{Current vs. voltage characteristic curves at different
temperatures for one of the studied lamellae (thickness $d = 800$~nm). The inset
blows out the data at low currents and low voltages. The
continuous lines were calculated following \cite{amb69} with the
critical Josephson current as the only free parameter, see text.}
\label{IV}
\end{figure}

In Fig.~\ref{VT} we recognize clear drops of the voltage at different
temperatures for lamellae with thickness $d \gtrsim 200~$nm, while
for a thinner lamella ($d = 100~$nm) a semiconducting-like
behavior is observed down to 2~K, the lowest measured temperature.
A similar semiconducting like dependence is measured for bulk graphite or multilayer
graphene samples without interfaces, independently of the lateral
size of the sample. For lamellae with thickness
$d \gtrsim 200$~nm the measured voltage at low enough temperatures
and currents fluctuates around zero within $\lesssim |\pm 20|~$nV.
The finite sensitivity of the electronics and the noise
 do not allow, strictly speaking,  to observe a zero
resistance state below  $T_c$. For that purpose one needs to show
the existence of persistent currents, as done in
\cite{kaw13}, for example. The current-voltage ($I-V$)
characteristic curves obtained for the studied lamellae are consistent  with the existence of Josephson
junctions, as Fig.~\ref{IV} shows for an 800~nm
thick lamella at different temperatures.

Thermal fluctuation effects dominate in a Josephson junction when
the thermal energy $k_B T$ is larger than the Josephson coupling
energy $E_J$ = $(\hbar/2e)I_c$ ($I_c$ is the critical Josephson
current). Early theoretical work provides an adequate framework to
understand the measured $I-V$ curves including the influence of
thermal fluctuations on Josephson junctions \cite{amb69,iva68}. It
is found that there is always a finite resistance, even below
$I_c$, due to thermally activated phase processes.   Using the differential equation proposed in
\cite{amb69}, we fit the measured  $I-V$ curves with  the critical
current $I_c(T)$ as the only free parameter. To present the
results in actual units of current and voltage  the values of the resistance in the
normal state $R_n$ have been used, obtained from the slopes of the
$I-V$ curves well above $I_c(T)$, once the linear regime was
reached. From the fits to the $I-V$ curves shown in Fig.~\ref{IV}
we obtained $I_c(\mu$A) = 0.162, 0.158, 0.091, and 0.013 at
$T($K$) = 40, 50, 70, 85~$K, respectively.  The rather unusual
behavior of the $I-V$ curves at large voltages, see Fig.~\ref{IV},
is  due to the transition from the metallic- to a
semiconducting-like temperature dependence at large currents,
similar to the magnetic-field driven metal-insulator transition
(MIT) in HOPG with interfaces \cite{kempa00}.
\begin{figure}[tbp]\centering
\includegraphics[width=0.95\columnwidth]{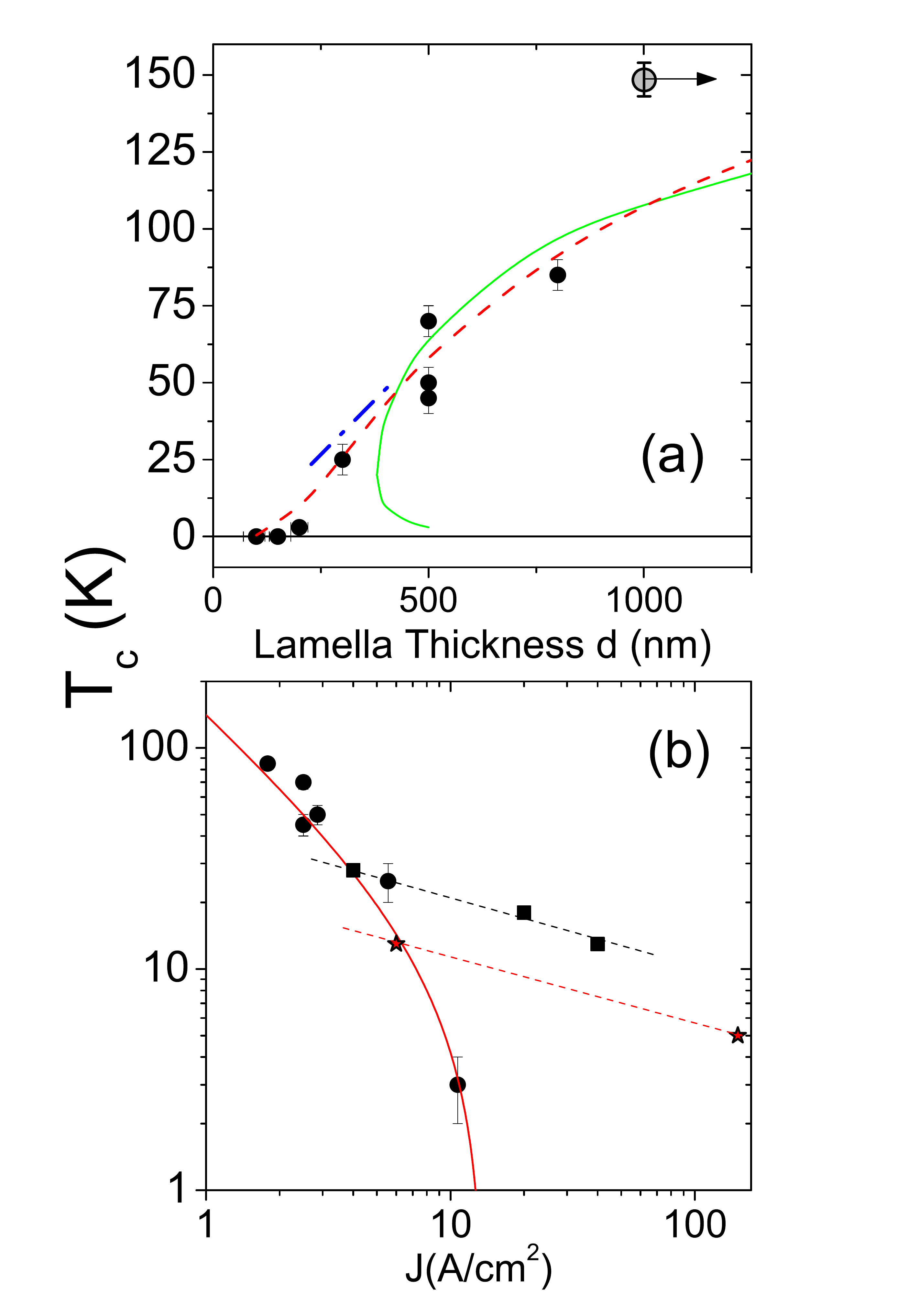}
\caption{ (a) Critical temperature $T_c$ defined at the onset of
the voltage decrease with decreasing temperature, obtained at 1~nA
input current vs. the thickness of the lamella. The straight
dashed-dotted (blue) line represents the linear fit through the
experimental points obtained for Nb/Al multilayers from
\cite{gui86} (both axis must be divided by 10 in this case). The
continuos (green) line follows Eq.(\ref{guimpelformula}) proposed in
\cite{gui86}. 
 The dashed (red) line
follows Eq.~(\ref{eq:fittedfunction}) with the parameters
 described in the text. The upper point represents
 the temperature of the maximum in the
 resistance for the bulk HOPG sample, see Fig.~\ref{VT}.
 The arrow indicates the expected range
 ($ 1~\mu$m $\lesssim d \lesssim 10~\mu$m) for the
 size of the interfaces in the bulk sample, taking into account
 TEM  \cite{bal13} as well as
 EBSD \cite{gon07} measurements. (b) The critical
 temperature vs. the estimated current density taking into account the
measured total normal area of the lamellae.
 The red line is a fit to the experimental values of $T_c$ obtained
  at 1~nA, which follows
the function $153(1/J)[$K cm$^2$/A]$ - 11$[K] $\propto d$. The
$T_c(J)$-curves of the two data sets ($\blacksquare, \star$,
dashed guide lines) were measured increasing
 the input current density in the same lamella.}
\label{Tcd}
\end{figure}

In  Fig.~\ref{VT} (right $y-$axis) we show the temperature
dependence of the resistance of a bulk piece of HOPG ZYA sample,
a part of the same bulk piece from where all the lamellae here studied were
obtained. The measured resistance shows a maximum at $T \simeq
150~$K. This maximum is not universal but
depends on the internal structure of the graphite sample.  Our
hypothesis is that the position of the maximum in the temperature
dependence of the resistance is directly related to the existence
of interfaces and their properties.
Following transport results in graphite flakes \cite{esqpip}
and an interpretation of the magnetic-field driven MIT in graphite
bulk samples \cite{kempa00}, we argue that the maximum in the electrical
resistance indicates the temperature below which a Josephson-like
coupling between the superconducting regions within the interfaces
starts to be measurable. In this case one possible reason for the
change in the temperature dependence of the resistance could be
the size of the superconducting 
regions or the superconducting/normal ratio.

Figure \ref{Tcd}(a) shows  $T_c$ obtained at 1~nA vs. the lamella
thickness $d$. We observe that the  defined $T_c$ decreases nearly
linearly with decreasing $d$ and extrapolates to zero at a finite thickness
$d \sim 160~$nm. To check whether this decrease of $T_c$ is not
due to an increase in the current density related to the change in the
lamellae geometry, we have measured also the current density
dependence of $T_c$ in two TEM lamellae. Figure \ref{Tcd}(b) shows
$T_c$ as a function of the current density changing the input
current for constant geometry. One sees clearly a much weaker
dependence than the one obtained changing the lamella thickness.

The  behavior of the granular superconductivity at graphite
interfaces is certainly not as simple as in usual
Josephson-coupled
 superconducting grains in a normal matrix. In real graphite
samples we expect inhomogeneously distributed superconducting
strength within the graphite interfaces due to, e.g., changes in
the stacking order, differences in the doping, etc. According to
\cite{mun13} high-temperature superconductivity at the interfaces
may survive throughout the sample due to the proximity effect
between ABC/ABA interfaces where the order parameter is enhanced.
To our knowledge there is no published theory directly applicable
to understand the interface-size dependence of the critical
behavior found in this work. Nevertheless, taking into account
that the observed behavior can be related to the existence of
superconducting and normal conducting regions, let us compare the
size dependence of $T_c$ obtained here with the one observed in
conventional superconductors. Especially interesting in this frame
is the linear decrease of the superconducting critical temperature
with the {\em whole} thickness of the ensemble of superconducting/metal
multilayers leaving {\em constant the thickness of each of the
layers}, see \cite{gui86} and refs. therein. The dashed-dotted
straight line in Fig.~\ref{Tcd}(a) is the experimental line
obtained for Nb/Al multilayers multiplying by 10 both axes. In this case
 the
obtained $T_c(d)$ dependence of our lamellae has a nearly identical slope as that
obtained for Nb/Al multilayers, where
the thickness is the total thickness  $d_T$ \cite{gui86}.

The origin for the change of $T_c$ in conventional superconducting
multilayers and thin wires has been tentatively given in
\cite{gui86,OF} based on weak localization (WL)  corrections to $T_c$
for 2D superconductors \cite{fuk84}. In both, the presence of disorder
affects the screening of the Coulomb interaction, and since the latter
changes the BCS coupling parameter, this results into an exponential
suppression of the critical temperature in ordinary BCS
superconductors. However, the parameter dependence of this exponential
is different in the theory \cite{OF} than the one used to fit the
experimental results of Nb/Al multilayers in \cite{gui86}. Below, we
compare our results to both approaches.

In \cite{OF}, the important aspect of the disorder correction is
screening, and therefore the relevant size scale is that of the whole conducting region, and the parameter controlling the correction is $t=(e^2/(2\pi^2 \hbar))R_\square$, where $R_\square$ is the sheet resistance \cite{OF}. In the limit $t \ll \gamma^2$, where $\gamma$ is the dimensionless bare BCS coupling parameter, the effective critical temperature can then be obtained as
\begin{equation}
T_c(d)=T_\infty \exp^{-t(d)/(6\gamma^3)} = T_\infty \exp^{-\alpha/d},
\label{eq:fittedfunction}
\end{equation}
where $\alpha=t(d)/(6 \gamma^3)$ and $T_\infty$ is the bulk critical temperature. The dashed line  in Fig.~\ref{Tcd} shows a fit of this type of behavior to our data. There, we used $T_\infty=200$ K and $\alpha \approx 600$ nm yielding the resistivity $\rho \simeq 0.05 \gamma^3$ $\Omega$m. Note that the theory in Eq.~(3) of Ref.~\cite{OF} also predicts that $T_c(d)=0$ if $t=2\gamma^2$. With our fitting parameters, this would take place for $d=200/\gamma$ nm. As we observe no superconducting response any more at $d \lesssim 160$ nm, this would indicate the presence of a very large $\gamma \approx 1$, which is strictly speaking outside the validity range of the approach in \cite{OF}. For this value of $\gamma$  the estimated resistivity $\rho$ is several orders
of magnitude larger than the one obtained from multilayer graphene
samples without interfaces \cite{dus11}.

On the other hand, \cite{gui86} compares the thickness to the thermal
length $\ell_{\rm th}(T)=(\hbar D/k_B T)^{1/2}$ at temperature $T$. Here, the 2D diffusion
constant $D = v_F \ell/2$, where $v_F \approx 10^6$ m/s is the Fermi
velocity and $\ell$ is the mean free path. Therefore, the estimated
correction to the critical temperature is
\begin{equation}
T_c(d)=T_\infty e^{-\ell_{\rm th}(T_c)/d}\,.
\label{guimpelformula}
\end{equation}
Independent measurements
done in graphite flakes without (or with much less influence of)
interfaces provide $\ell \sim 3$ $\mu$m at $T<100$ K \cite{esq12} and
therefore $D \sim 1.5\times 10^4$ cm$^2$/s. Note that this is four
orders of magnitude larger than the one used in
\cite{gui86}, meaning that the effect is relevant in far thicker
samples or at much higher temperatures than in Nb/Al multilayers which
have $T_c \lesssim 10$ K. We use this diffusion constant and for
$T_\infty = 150~$K  we take the temperature of the maximum resistance measured
in HOPG (see Fig.~\ref{VT}). The obtained numerical solution of
Eq.~\eqref{guimpelformula} is plotted as the continuous line in
Fig.~\ref{Tcd}. The semiquantitative agreement is
remarkable as well as the estimated cut-off $d_{\rm min} = \ell_{\rm
  th}(T_c)/2.7 \approx 0.38$ $\mu$m, below which
Eq.~\eqref{guimpelformula} has no solution.

We note that using either the theory of \cite{OF} or the model used in
\cite{gui86} to describe our results has
many shortcomings: For interface superconductivity, we expect
screening to be strongly inhomogeneous, whereas \cite{OF} assumes
homogeneous superconductors and in \cite{gui86} the metals did not
differ very much. Moreover, even the use of the conventional weak-coupling theory
in describing our results is questionable, because of the high
critical temperatures. One possible explanation of this high critical
temperature is associated with the flat bands emerging at the interfaces due to inclusions with rhombohedral stacking \cite{kop11,kop13}. However, a prediction of the screening effect on such flat band superconductivity does not yet exist.

Finally, it should be noted that some evidence for the formation of
charge density waves (CDW) was found in CaC$_6$ at $\sim 250~$K, whereas
its $T_c = 11.5~$K \cite{rah11}. On the other hand doping the surface of bulk
graphite samples with Ca atoms showed hints for
superconductivity at $T \sim 200~$K \cite{han10}.
Taking into account the available evidence on the antagonist
relationship between  CDW and superconductivity \cite{gab13}, future
experiments should try to check whether CDW are also formed in
HOPG interfaces and their possible relation with the size effect reported in this study.

In conclusion, we have found that the temperature at which the
Josephson behavior in TEM lamellae sets in, decreases with the
size of the interfaces. This behavior provides a way to understand
differences in the temperature dependence of the resistance in
graphite samples with interfaces. Weak localization effects appear
to be a possible origin for the here reported phenomenon.


This work was partially supported by the ESF ``Energie", the European Research Council (Grant No. 240362-Heattronics)
and the Academy of Finland.




\end{document}